\documentclass[paper,noblind]{geophysics}
\usepackage{graphicx}
\usepackage{caption}

\begin{document}

\title{Target-oriented least-squares reverse-time migration with Marchenko redatuming and double-focusing: Field data application}

\renewcommand{\thefootnote}{\fnsymbol{footnote}} 

\ms{GEO-2023-0374.R1} 

\address{
\footnotemark[1]Department of Geoscience and Engineering, Delft University of Technology, PO Box 5048, 2600 GA Delft, The Netherlands \\
\footnotemark[2]Department of Imaging Physics, Delft University of Technology, PO Box 5048, 2600 GA Delft, The Netherlands}
\author{Aydin Shoja\footnotemark[1], Joost van der Neut\footnotemark[2] and Kees Wapenaar\footnotemark[1]}

\footer{}
\lefthead{Shoja et al.}
\righthead{Norwegian Sea target-oriented LSRTM}

\maketitle

\begin{abstract}
Recently, the focus of reflection seismologists has shifted to applications where a high-resolution image of the subsurface is required. Least-Squares Reverse-Time Migration (LSRTM) is a common tool used to compute such images. Still, its high computational costs have led seismologists to use target-oriented LSRTM for imaging only a small target of interest within a larger subsurface block. Redatuming the data to the upper boundary of the target of interest is one approach to target-oriented LSRTM. Still, many redatuming methods cannot account for multiple scattering within the overburden. We apply a target-oriented least-squares reverse time migration algorithm that integrates Marchenko redatuming and double-focusing to a field dataset. This redatuming method accounts for all orders of multiple scattering in the overburden, thus improving the accuracy of target-oriented LSRTM. Moreover, we demonstrate the effectiveness of a double-focusing algorithm in reducing the data size by decreasing both spatial and temporal dimensions of the model and the data. The algorithm's performance is evaluated using field data acquired in the Norwegian Sea. The numerical results show that our target-oriented LSRTM algorithm can reduce the internal multiple effects and increase the resolution of the resulting image.


\end{abstract}

\section{Introduction}
Seismic imaging and inversion are a set of techniques used by geophysicists to estimate parameters related to wave propagation, such as reflectivity, velocity, and density, within the Earth's subsurface. A network of sources and receivers is positioned on the Earth's surface to produce and record seismic waves, from which these parameters are determined. Geophysicists typically assume a subsurface model that consists of a background model ($m_0$) for long wavelengths and a perturbation model for short wavelengths ($\delta m$), based on a weak-scattering assumption \citep{SeisInv,Claerbout}. The primary objective of seismic imaging is to generate a structural image of the short-wavelength perturbation model ($\delta m$).

Reverse-Time Migration (RTM) is a popular method among different imaging techniques since it can produce high-resolution images and better handle complex geological structures \citep{BaysalRTM,McMechan,Zhou,Zhang}. RTM creates images by cross-correlating the forward-propagated wavefield and its back-propagated counterpart based on the Born approximation. However, improving the resolution and quality of RTM images is possible by inverting the Lippmann-Schwinger integral under the Born approximation for the perturbation model with a least-squares algorithm \citep{Dutta,Tang,Liu,Huang2,Kaur}. This inversion process is known as Least-Squares Reverse-time migration (LSRTM).

However, LSRTM is a computationally expensive algorithm \citep{Dai,Tang,Herrmann,Farshad}. To reduce the computational cost of LSRTM, one can restrict the model's dimensions by focusing on a small area inside the big block of the subsurface model. To compute the image of this smaller region, the wavefield on the upper boundary of this region is needed at least. The process of computing the wavefield on the boundary of this target from surface recorded data is called redatuming \citep{Valenciano,Haffinger,Willemsen,Yuan,Zhao,Guo2,Ravasi2,Huang1,EBoindi}. One prominent redatuming technique is Marchenko redatuming \citep{Wapenaar6,Wapenaar7,Diekman2}.

Marchenko redatuming \citep{JoostGJI2015,Wapenaar5,Wapenaar6,Wapenaar7,JoostScaling,Dukalski} can create virtual receivers on the boundary of the target of interest while accounting for all orders of internal multiple scattering effects and reflections. Since Marchenko redatuming and Green's functions retrieval are powerful tools, researchers use them to address seismic imaging and inversion issues \citep{TCui,Cui,Zhang,Diekmann3}. Moreover, it is possible to perform a double-sided redatuming using Marchenko focusing functions. Double-sided redatuming creates virtual sources in addition to virtual receivers at the boundary of the target. The process of double-sided redatuming is also called double-focusing  \citep{Staring,Shoja3}. Marchenko double-focused wavefields account for all orders of internal multiples generated inside the overburden, enabling us to create images with less impact from internal multiples. Moreover, Marchenko double-focusing compacts the data's time axis, reducing its size even more.

This paper combines the Marchenko double-focusing and target-oriented LSRTM algorithm to create high-resolution artifact-free images of a marine data set from the Vøring region in the Norwegian Sea. First, we review the theory of target-oriented LSRTM with Marchenko double-focusing, which is fully developed and is validated with synthetic models by \cite{Shoja3}. Second, we apply this algorithm to a marine dataset, and finally, we discuss the results and conclude the paper. 
\section{Theory} \label{sec:2}

\subsection{Least-squares reverse-time migration}

\cite{Dai} show that classical RTM can be derived from the Born approximation of seismic reflection data. In the Born approximation, the incident wavefield ($P^{inc}$) can be estimated using the background Green's function. The perturbation model is expressed as $\delta m=(\frac{1}{c^2}-\frac{1}{c_0^2})$ where $c$ represents the medium velocity and $c_0$ represents the background velocity. This equation links $\delta m$ to the scattered data in the frequency domain ($P^{scat}$) through a linear relation \citep{born_wolf_1999,SeisInv,Peter}:
\begin{equation}
 P^{scat}_{pred}(\textbf{x}_r,\textbf{x}_s,\delta m,\omega) = \frac{\omega^2}{\rho_0} \int_{V} G_0(\textbf{x}_r,\textbf{x},\omega)\delta m(\textbf{x})P^{inc}(\textbf{x},\textbf{x}_s,\omega),d\textbf{x}.
 \label{scat_int}
\end{equation}
The integral in Equation~\ref{scat_int} is computed throughout the model's volume ($V$). Here,
$P^{inc}(\textbf{x},\textbf{x}_s,\omega) = G_0(\textbf{x},\textbf{x}_s,\omega)W(\omega)$.
 Moreover, $\omega$ is the angular frequency, $W$ is the source signature, $G_0$ is the Green's function computed in the background model ($c_0$), $\rho_0$ is the background density, and $P^{scat}_{pred}$ is the scattered predicted data. The subscripts "$r$" and "$s$" indicate the receiver and source, respectively. This equation can be expressed in an operator format as follows:
\begin{equation}
 P^{scat}_{pred}(\textbf{x}_r,\textbf{x}_s,\delta m,\omega) = \mathcal{L}\delta m.
 \label{scat}
\end{equation}
Here $\mathcal{L}$ is the forward Born operator.

The standard method of reverse-time migration involves obtaining an approximate perturbation model by taking the adjoint of $\mathcal{L}$ and applying it to the observed scattered data:
\begin{equation}
    \delta m^{mig}(\textbf{x}) = \mathcal{L}^\dagger P^{scat}_{obs}.
\end{equation}
Since the adjoint of this kernel is merely an approximation of its inverse, the resolution of the perturbation model obtained through this process is limited.

To tackle the problem of limited resolution, scholars have adopted a least-squares strategy in which the adjoint operator ($\mathcal{L}^\dagger$) is substituted with a damped least-squares solution  \citep{Marquardt,Dai,Dutta}:
\begin{equation}
    \delta m^{mig} = [\mathcal{L}^\dagger \mathcal{L}+\epsilon]^{-1}\mathcal{L}^\dagger P^{scat}_{obs}.
\end{equation}
Here, $\mathcal{L}^{\dagger}\mathcal{L}$ is the Hessian matrix, and $\epsilon$ is a damping factor.Unfortunately, calculating the Hessian matrix ($\mathcal{L}^{\dagger}\mathcal{L}$) and its inverse is computationally infeasible. As an alternative, an iterative algorithm that minimizes the L2-norm of the discrepancy between the observed and predicted data is often used to update the perturbation model:
\begin{equation}
    C(\delta m) = \frac{1}{2}\big\| P^{scat}_{pred}(\delta m)-P^{scat}_{obs}\big\|_2^2.
    \label{norm2}
\end{equation}
One potential way to tackle this optimization problem is by utilizing a conjugate gradient algorithm \citep{nocedal_wright_2006}. In least-squares reverse-time migration, the background velocity model ($c_0(\textbf{x}$)) is not changed, and only the perturbation model ($\delta m$) is updated, resulting in the Green's functions of Equation~\ref{scat_int} being calculated only once. To learn more about least-squares reverse-time migration, please see \cite{SeisInv}.

\subsection{Marchenko redatuming and double-focusing}
Marchenko redatuming is an innovative data-driven technique that can recover the Green's function generated by a source at the surface and recorded by a virtual receiver just above the target area's surface, including all orders of multiple-scattered events. This method only requires the reflection response at the surface and a smooth background velocity model of the overburden capable of predicting the direct arrival from the surface to the redatuming level.

The following coupled Marchenko-type representations are solved iteratively to retrieve the Green's functions at the redatuming level \citep{Wapenaar6}:
\begin{equation}
    G^-_{Mar}(\textbf{x}_{v},\textbf{x}_{r},\omega) = \int_{\mathcal{D}_{acq}} R(\textbf{x}_r,\textbf{x}_s,\omega)f_1^+(\textbf{x}_s,\textbf{x}_{v},\omega) \,d\textbf{x}_s
    -f_1^-(\textbf{x}_{r},\textbf{x}_{v},\omega), 
    \label{Mar_r-}
\end{equation}
and
\begin{equation}
    G^+_{Mar}(\textbf{x}_{v},\textbf{x}_{r},\omega) = -\int_{\mathcal{D}_{acq}} R(\textbf{x}_r,\textbf{x}_s,\omega)f_1^-(\textbf{x}_s,\textbf{x}_{v},\omega)^* \,d\textbf{x}_s
    + f_1^+(\textbf{x}_{r},\textbf{x}_{v},\omega)^* .
    \label{Mar_r+}
\end{equation}
In these equations, $\mathcal{D}_{acq}$ represents the acquisition surface where $\textbf{x}_s$ and $\textbf{x}_r$ are situated. $G^-_{Mar}$ and $G^+_{Mar}$ denote the up-going and down-going components of the Marchenko redatumed Green's function, respectively (see Fig.~\ref{fig:Figure1a.eps} and \ref{fig:Figure1b.eps}). Additionally, $f_{1}^-(\textbf{x}_s,\textbf{x}_{v},\omega)$ and $f_{1}^+(\textbf{x}_s,\textbf{x}_{v},\omega)$ denote the up-going and down-going parts of the focusing function, respectively, with the subscript "$v$" denoting a virtual point situated on the redatuming level denoted by $\mathcal{D}_{tar}$. Furthermore, $R(\textbf{x}_r,\textbf{x}_s,\omega)$ refers to the dipole response of the medium at the acquisition surface, and it is related to the up-going Green's function at the acquisition surface ($G^-$) via the following relationship \citep{Kees89}:
\begin{equation}
    R(\textbf{x}_r,\textbf{x}_s,\omega) = \frac{\partial_{3,s} G^-(\textbf{x}_r,\textbf{x}_s,\omega)}{\frac{1}{2}i\omega\rho(\textbf{x}_s)}.
\label{R_to_r}
\end{equation}
The partial derivative in the downward direction taken at $\textbf{x}_s$ is denoted by $\partial_{3,s}$. This partial vertical derivative is computed in the frequency-wavenumber domain by multiplying the wavefield by $ik_z$, where $k_z$ is the vertical wavenumber. $\rho(\textbf{x}_s)$ is the density at $\textbf{x}_s$. It is important to remove horizontally propagating waves and surface-related multiples before inserting $R(\textbf{x}_r,\textbf{x}_s,\omega)$ into Equations~\ref{Mar_r-} and \ref{Mar_r+}. The detailed derivation of these integrals and their solution for computing the focusing functions and Green's functions can be found in \cite{Wapenaar6} and \cite{Thorbecke2}.

The above-mentioned equations correspond to single-sided redatuming. To perform a double-sided redatuming, a convolution operation on the up-going and down-going parts of the Marchenko redatumed Green's function is proposed by \cite{Staring}. This operation involves filtering the down-going focusing function in a multi-dimensional manner:
\begin{equation}
    G^{-,+}_{df}(\textbf{x}_{v},\textbf{x}'_{v},\omega) = \int_{\mathcal{D}_{acq}}  G^-_{Mar}(\textbf{x}_{v},\textbf{x}_{r},\omega)\mathcal{F}_1^+(\textbf{x}_r,\textbf{x}'_{v},\omega)\,d\textbf{x}_r,
\label{double-G-}
\end{equation}
and
\begin{equation}
    G^{+,+}_{df}(\textbf{x}_{v},\textbf{x}'_{v},\omega) = \int_{\mathcal{D}_{acq}} G^+_{Mar}(\textbf{x}_{v},\textbf{x}_{r},\omega)
    \mathcal{F}_1^+(\textbf{x}_r,\textbf{x}'_{v},\omega)\,d\textbf{x}_r,
\label{double-G+}
\end{equation}
where
\begin{equation}
    \mathcal{F}_1^+(\textbf{x}_r,\textbf{x}'_{v},\omega) = \frac{\partial_{3,r} f_1^+(\textbf{x}_r,\textbf{x}'_{v},\omega)}{\frac{1}{2}i\omega\rho(\textbf{x}_r)}.
\end{equation}
Here the vertical derivative is taken with respect to $\textbf{x}_r$, which is computed similarly to $\partial_{3,s}$, as discussed below Equation \ref{R_to_r}.
Equations~\ref{double-G-} and~\ref{double-G+} use superscripts to indicate the direction of propagation at the receiver and source locations, respectively. The term "$df$" stands for "double-focused." This process is referred to as "Marchenko double-focusing."

The Marchenko double-focusing technique yields two Green's functions, namely a down-going ($G^{+,+}_{df}$) and an up-going ($G^{-,+}_{df}$) Green's function (fig.~\ref{fig:Figure1c.eps} and \ref{fig:Figure1d.eps}). The down-going Green's function consists of a band-limited delta function and interactions between the target and the overburden. $G^{-,+}_{df}$ can be interpreted as the continuation of propagation of $G^{+,+}_{df}$ through the target and recording the up-going part of it at the redatuming level. This up-going wavefield includes interactions between the target and the overburden on the source side. In contrast, the conventional double-focusing approach involves using the inverse of the direct arrival of the transmission response of the overburden instead of the down-going Marchenko focusing function. However, this approach cannot predict and remove the multiples generated by the overburden. In subsequent sections, the term "double-focusing" is a general expression that refers to both approaches, and it is explicitly mentioned where a distinction between the methods is necessary.



\multiplot*{4}{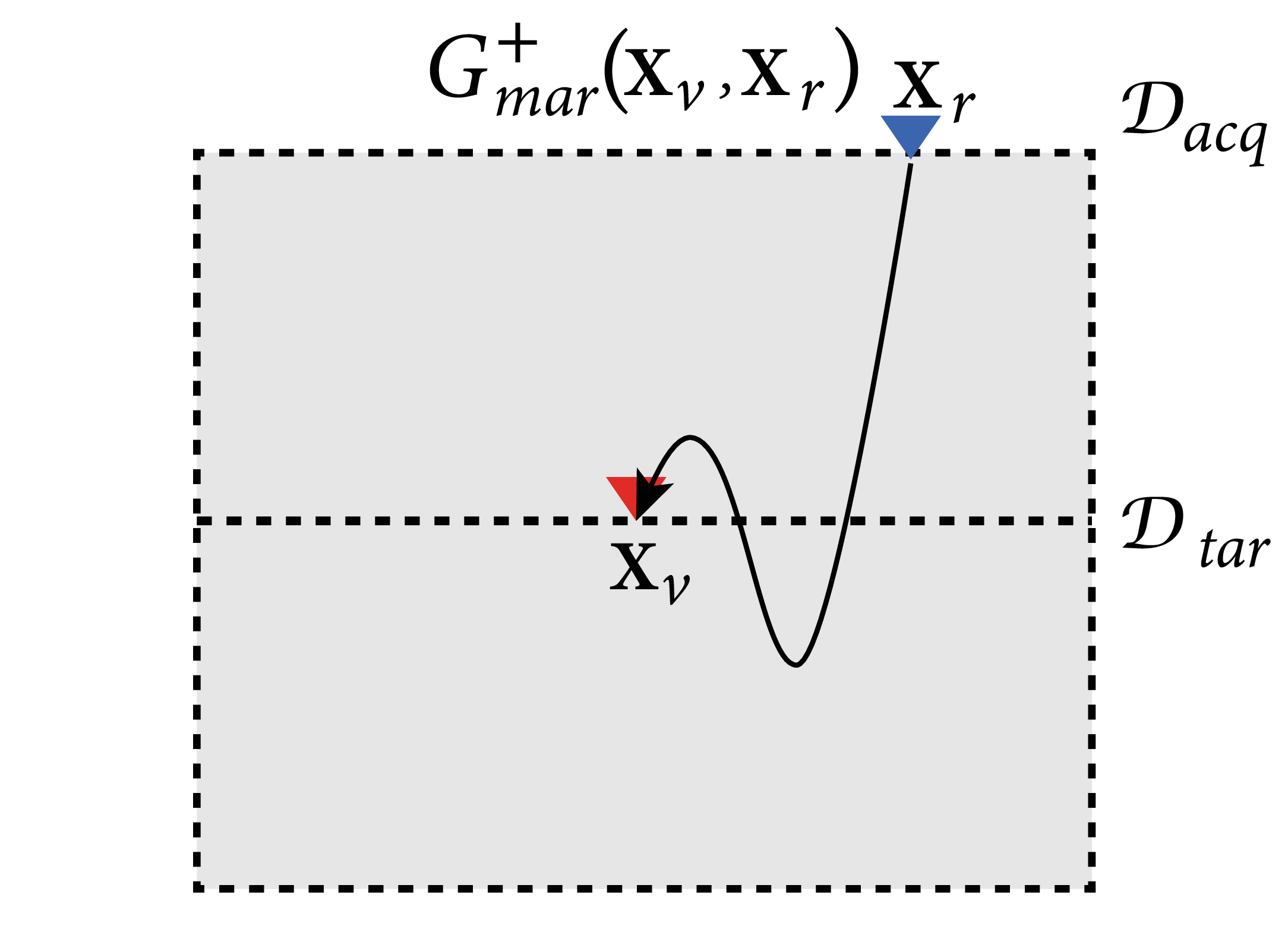,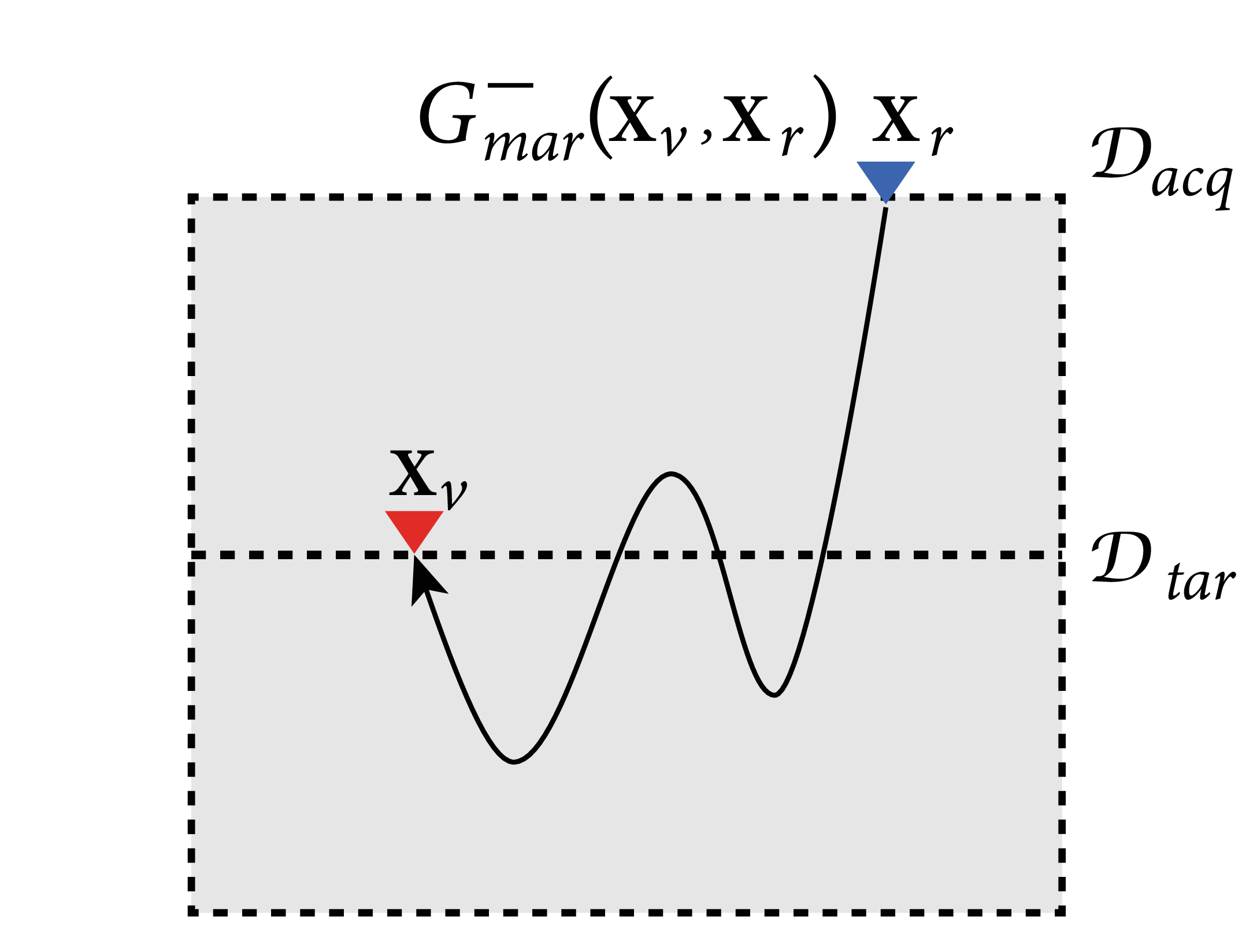,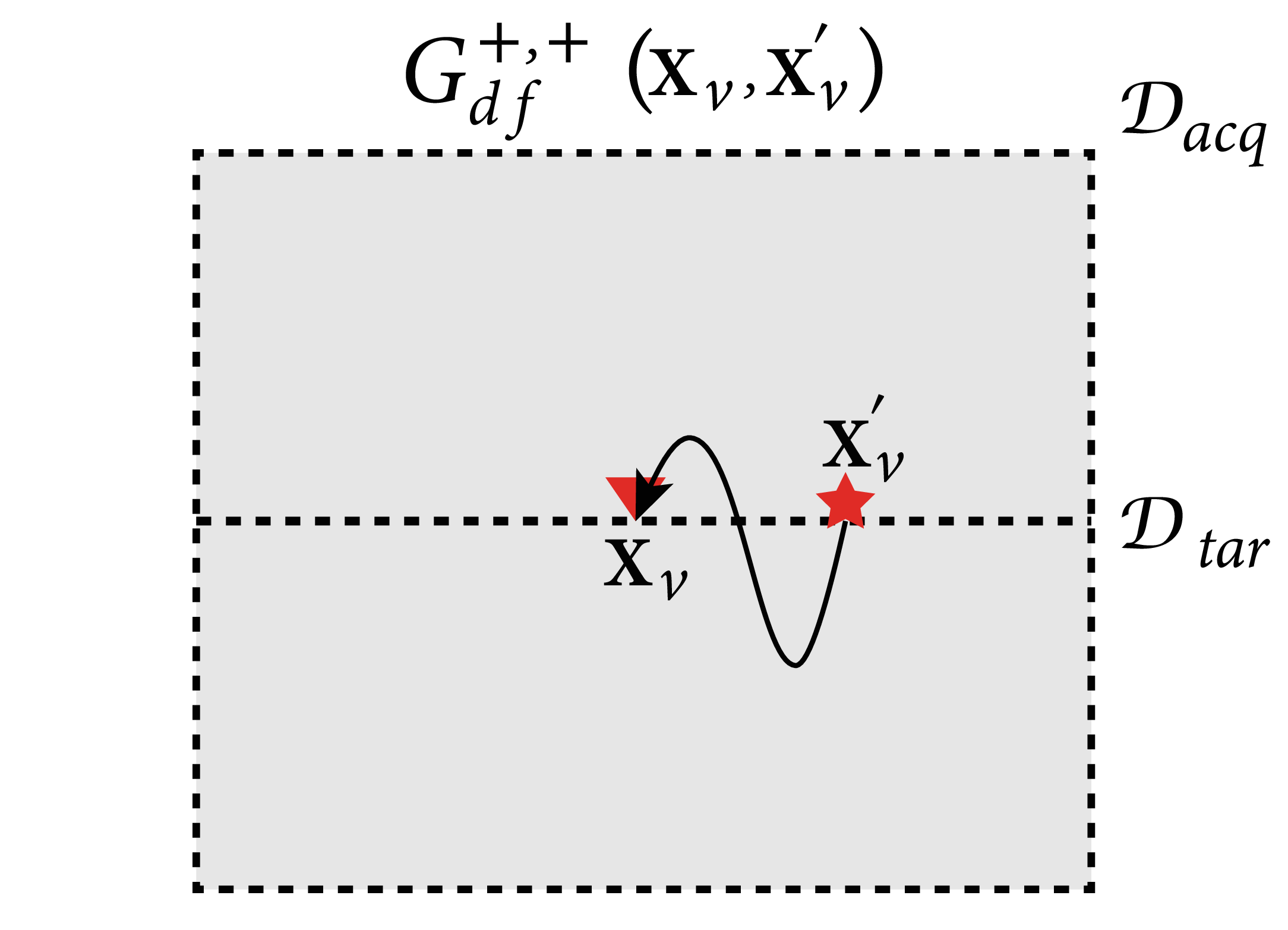,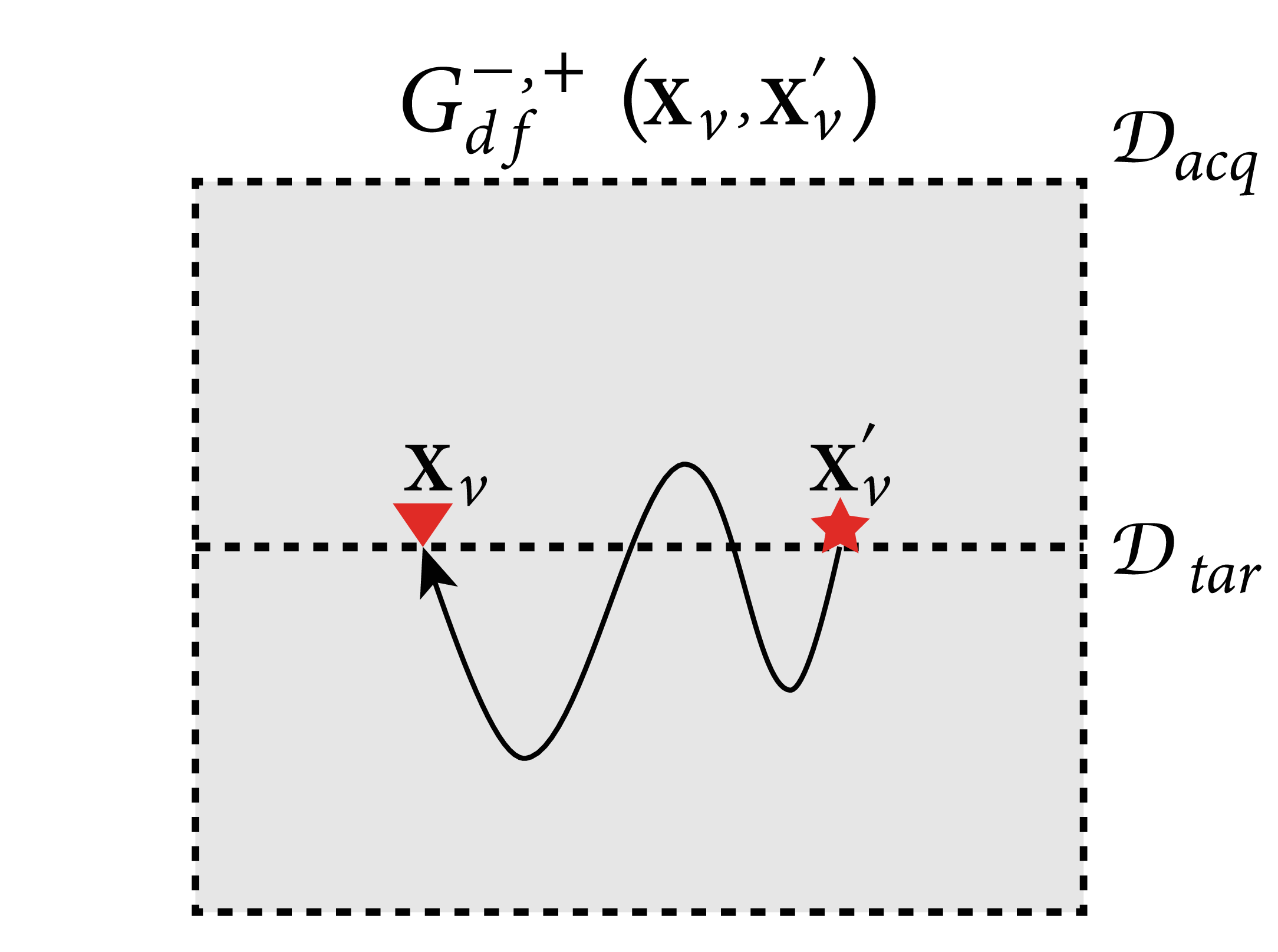}{width=0.4\columnwidth}
{The Green's functions resulting from Marchenko redatuming and double-focusing. a) down-going part of Marchenko Green's function, b) up-going part of Marchenko Green's function, c) down-going Marchenko double-focused Green's function, and d) up-going Marchenko double-focused Green's function.}

\subsection{Target-oriented LSRTM by Marchenko double-focusing}
The theory of the method is fully described in \cite{Shoja3}. Here, we give a brief explanation of the theory. The following integral is the base for target-oriented LSRTM by Marchenko double-focusing:

\begin{equation}
    \hat{P}^{scat}_{pred}(\textbf{x}'_{vr},\textbf{x}'_{vs},\delta m,\omega) = \frac{\omega^2}{\rho_0} \int_{\cal V} \hat{G}_0(\textbf{x}'_{vr},\textbf{x},\omega) \delta m(\textbf{x}) P^{inc}_{df}(\textbf{x},\textbf{x}'_{vs},\omega) d\textbf{x}.
    \label{P_pred2}
\end{equation}
Here, $\cal V$ is the target volume, $\textbf{x}$ is a point inside the target, and $\textbf{x}'_{vs}$ and $\textbf{x}'_{vr}$ are the virtual source and virtual receiver locations on the upper boundary of the target, respectively. Moreover,
\begin{equation}
    P^{inc}_{df}(\textbf{x},\textbf{x}'_{vs},\omega) = \int_{\mathcal{D}_{tar}}\frac{\partial_{3,vs}G_0(\textbf{x},\textbf{x}_{vs},\omega)}{\frac{1}{2}i\omega\rho(\textbf{x}_{vs})} G^{+,+}_{df}(\textbf{x}_{vs},\textbf{x}'_{vs},\omega)W(\omega)\,d\textbf{x}_{vs},
    \label{Pincdf}
\end{equation}
and
\begin{equation}
    \hat{G}_0(\textbf{x}'_{vr},\textbf{x},\omega) = \int_{\mathcal{D}_{tar}}  \Gamma(\textbf{x}'_{vr},\textbf{x}_{vr},\omega) G_0(\textbf{x}_{vr},\textbf{x},\omega)\,d\textbf{x}_{vr},
\end{equation}
where
 \begin{equation}
    \Gamma(\textbf{x}'_{vr},\textbf{x}_{vr},\omega) = \int_{\mathcal{D}_{acq}} G_d^+(\textbf{x}'_{vr},\textbf{x}_s,\omega)^{-1} G_d^+(\textbf{x}_{vr},\textbf{x}_s,\omega)\,d\textbf{x}_s
 \label{PSF}
\end{equation}
is a point-spread function that acts as a band limitation filter on the predicted data. In Equation \ref{PSF}, $G_d^+$ is the first arrival of the Green's function between the target boundary and the surface. For a complete derivation of the above equations and an analysis of the effects of the point-spread function ($\Gamma(\textbf{x}'_{vr},\textbf{x}_{vr},\omega)$), we refer to \cite{Shoja3}.
Thus, the new cost function is:
\begin{equation}
    C(\delta m) = \frac{1}{2}\big\| \hat{P}^{scat}_{pred}(\delta m)-\hat{P}^{scat}_{obs}\big\|_2^2,
    \label{Ctar}
\end{equation}
where
\begin{equation}
    \hat{P}^{scat}_{obs} =G^{-,+}_{df}(\textbf{x}'_{vr},\textbf{x}'_{vs},\omega)W(\omega).
\end{equation}
We solve Equation \ref{Ctar} with a conjugate gradient algorithm.
\section{Field data example}
In a previous paper \citep{Shoja3}, target-oriented LSRTM with double-focusing is tested on synthetic models. Here, we apply this method to a field data set.
\subsection{Field data explanation}
\begin{figure}
    \centering
    \includegraphics[width=1\textwidth]{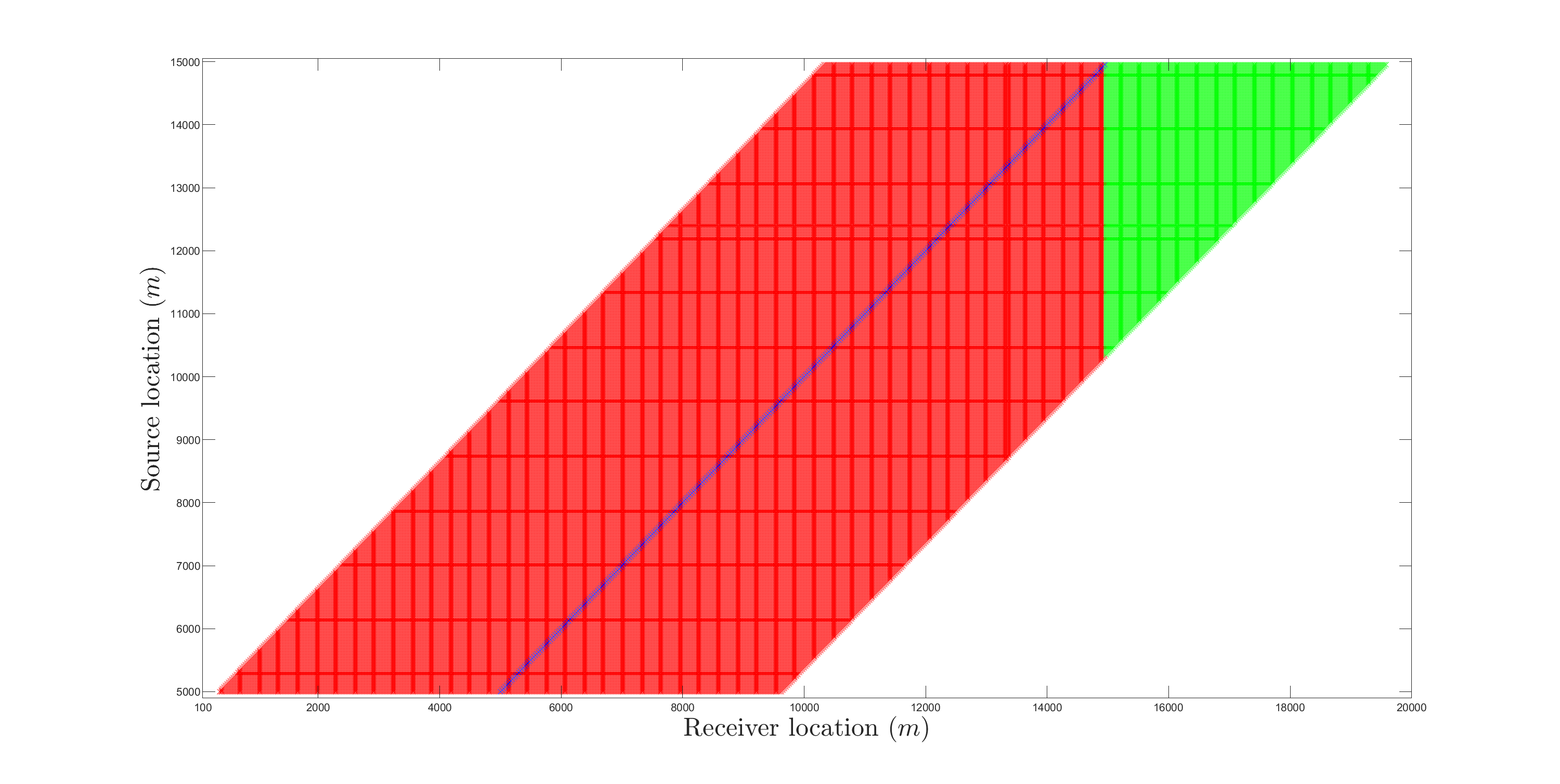}
    \caption{The acquisition geometry of the data set. Blue crosses show the source locations, red crosses show the receivers' locations and green crosses show the dummy traces added after source-receiver reciprocity to have an equal number of receivers per shot. The receivers on the left side of the sources are the real ones, and the receivers on the right side are added by source-receiver reciprocity.}
    \label{fig:2}
\end{figure}

\begin{figure}
    \centering
    \includegraphics[width=1\textwidth]{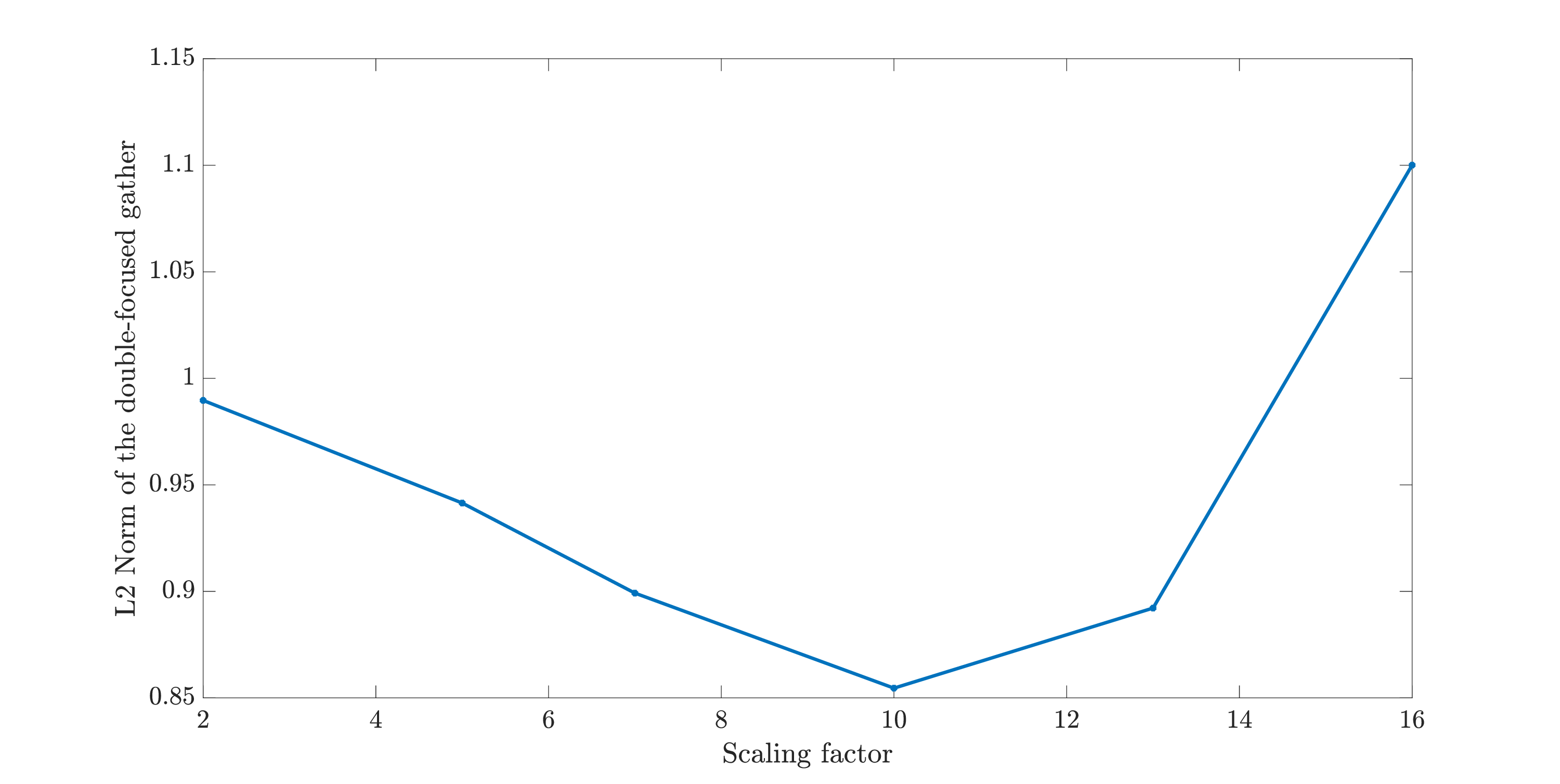}
    \caption{L2 norm of the gather shown in Figure 6a against different scaling values.}
    \label{fig:3}
\end{figure}

This part of the paper shows the results of the Marchenko-based target-oriented LSRTM on a field dataset provided by Equinor, which was acquired in the Norwegian Sea in 1994. The water bottom depth is $1.5$ km, which is deep enough to separate the free-surface multiple reflections from the primary and internal multiple reflections. The field dataset contains 399 shot gathers with 180 traces per gather, and the spatial sampling of sources and receivers is $25$ m. The field dataset was processed using the method proposed by \cite{EricDavy}, which involved muting the direct wave, estimating near-offset traces through the parabolic Radon transform \citep{Kabir}, compensating for 3D effects by multiplying with $\sqrt{t}$, and deconvolving the source wavelet. 
Source-receiver reciprocity is also applied to create offsets in the positive direction to prepare the dataset for the Estimation of Primaries through Sparse Inversion (EPSI) method to remove free-surface multiples \citep{vanGroenestijn}. After source-receiver reciprocity, each gather contains 371 receivers. Since it is not possible to recover the traces after the last shot with source-receiver reciprocity, dummy traces are added in this part of the data to have an equal number of traces in each shot gather. Table~\ref{tbl:1} shows the acquisition parameters, and Figure~\ref{fig:2} shows the acquisition geometry of this dataset. 
Adding reciprocal traces is also in line with having regularly sampled data for the Marchenko scheme. \cite{JohnoIrr} study the possibility of using data with irregular sampling. 
We apply a time gain of $1.73e^{1.3t}$ to the reflection response as recommended by \cite{JoeriSAGA} to compensate for the absorption effect. However, with this scaling function, the Marchenko redatuming procedure does not sufficiently reduce the multiple reflections energy for imaging. An incorrect scaling function can result in more artifacts \citep{JoostGJI2015}. Thus, we multiplied the reflection response already scaled with the aforementioned scaling function, using a range of values to adjust it for imaging. Then, we measured the L2-norm of the double-focused gather to find the value that produces the minimum energy \citep{JoostScaling,JoeriMasters}. Figure~\ref{fig:3} shows the L2 norm of the double-focused gather against the values we use. According to Figure~\ref{fig:3}, we choose value 10, which results in an adjusted scaling factor of $17.3e^{1.3t}$ for the original (non-scaled) reflection response.  \\
\\

\begin{table}[h!]
\centering
\begin{tabular}{|c|c|}
\hline
Parameter & Value \\
\hline
Number of source positions & 399 \\
Source spacing & 25 m \\
First source position & 5,000 m \\
Final source position & 14,950 m \\
Number of receiver positions per source & 180 \\
Receiver spacing & 25 m \\
Minimum source-receiver offset & 150 m \\
Maximum source-receiver offset & 4,625 m \\
Number of time samples & 2001 \\
Sampling rate & 0.004 s \\
High-cut frequency & 90 Hz \\
\hline
\end{tabular}
\caption{Acquisition Parameters for the dataset}
\label{tbl:1}
\end{table}

\begin{figure}
    \centering
    \includegraphics[width=0.7\textwidth]{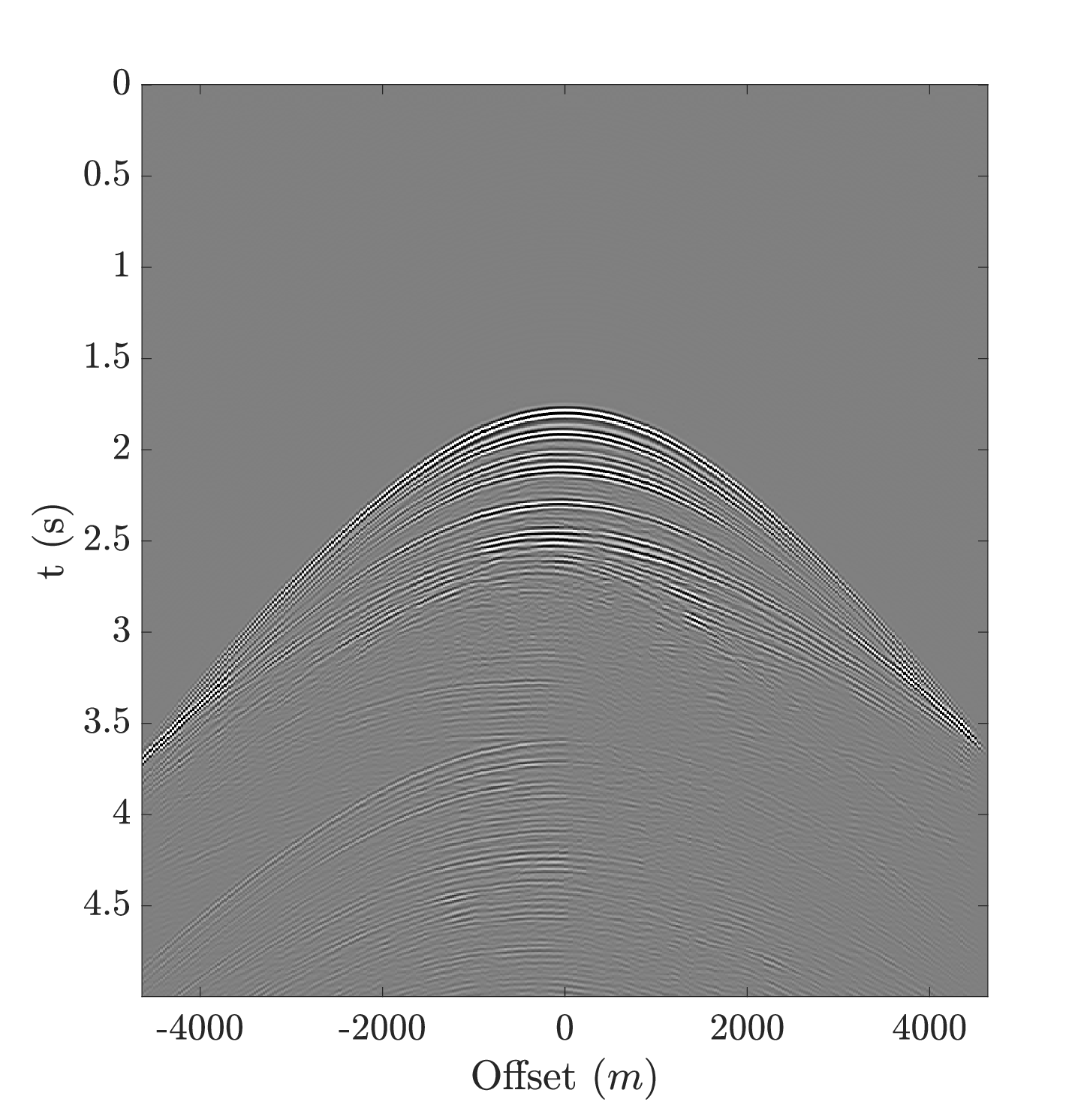}
    \caption{Reflection response with a source located at $\textbf{x}_s=$ (5000 m, 0 m). A Ricker wavelet with a dominant frequency of $30$ Hz is convolved with the data for better visualization.}
    \label{fig:4}
\end{figure}

Figure~\ref{fig:4} shows the surface reflection response after preprocessing, with a source located at $\textbf{x}_s=$ (5000 m, 0 m). We choose two different targets inside the medium. 

\subsection{LSRTM with double-focusing}
\subsubsection{Target of interest 1}

\begin{figure}
    \centering
    \includegraphics[width=1\textwidth]{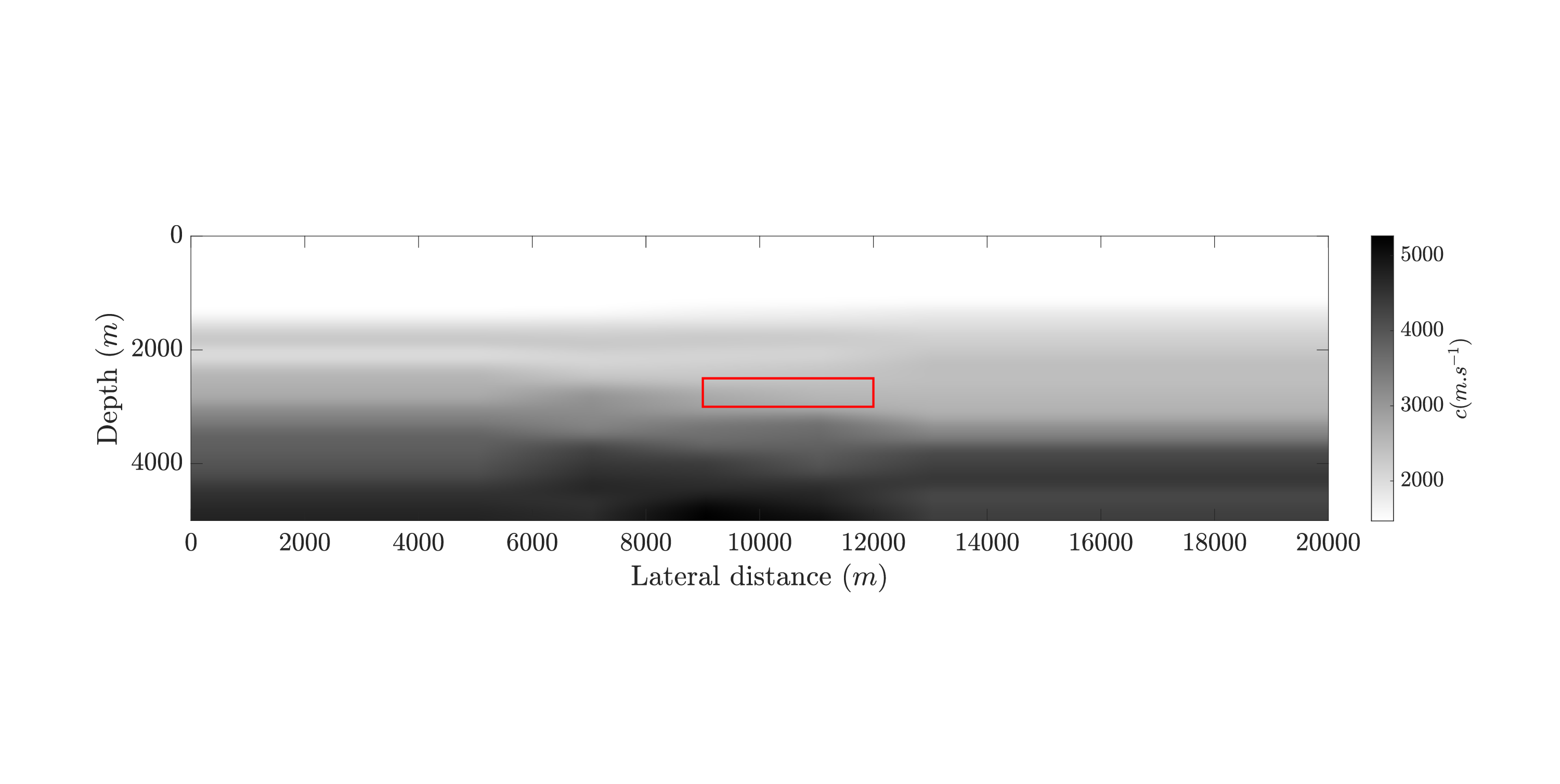}
    \caption{The smooth velocity model provided by Equinor for migration. The red rectangle inside the velocity model indicates the first target area. The virtual sources and receivers' positions are at the upper boundary of this target area.}
    \label{fig:5}
\end{figure}

Figure~\ref{fig:5} shows the smooth velocity model provided by Equinor for migration. The red rectangle inside the velocity model indicates the target area and the virtual sources and receivers' positions are at the upper boundary of this target area. We assume a constant density model for this study.

\begin{figure}
    \centering
    \includegraphics[width=1\textwidth]{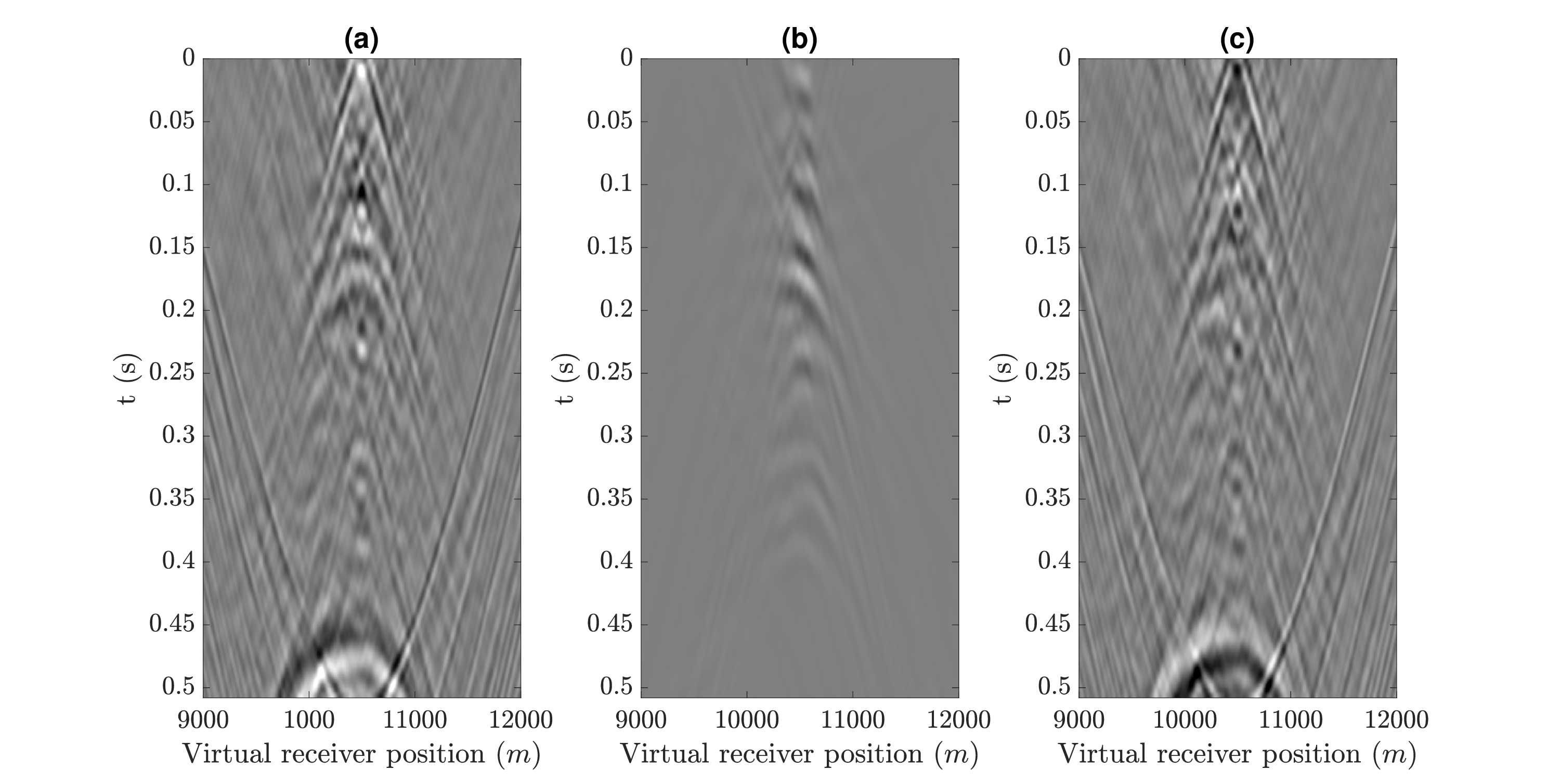}
    \caption{Marchenko double-focused data with a virtual source located at $\textbf{x}_{vs}=$ (10500 m, 2500 m) and virtual receivers at the same depth as virtual sources. a) 'observed data' obtained by Marchenko double focusing, b) predicted data after 35 iterations of LSRTM, and c) residuals after 35 iterations of LSRTM.}
    \label{fig:6}
\end{figure}

\begin{figure}
    \centering
    \includegraphics[width=1\textwidth]{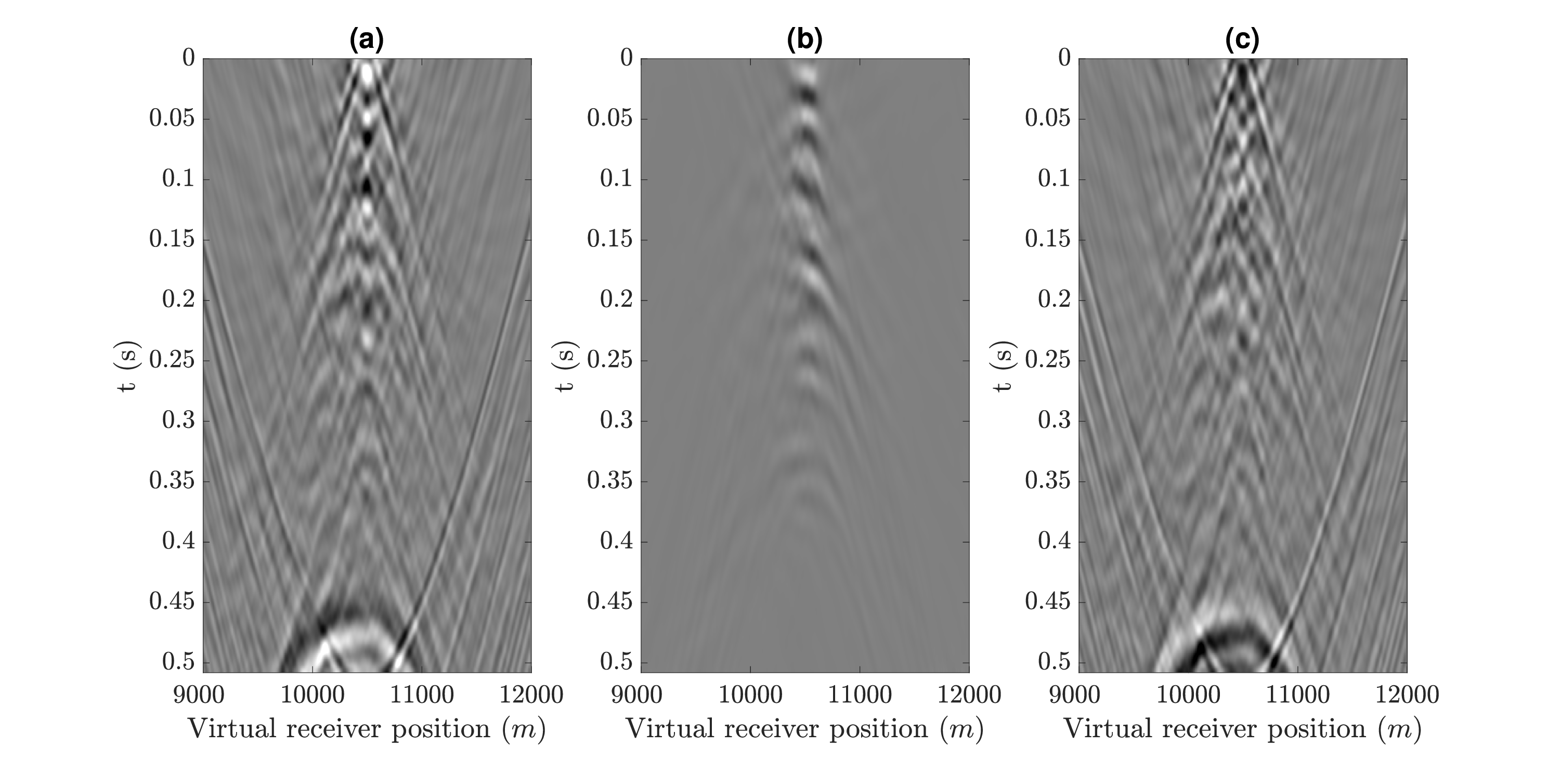}
    \caption{Conventional double-focused data with a virtual source located at $\textbf{x}_{vs}=$ (10500 m, 2500 m). a) 'observed data' obtained by conventional double-focusing, b) predicted data after 35 iterations of LSRTM, and c) residuals after 35 iterations of LSRTM.}
    \label{fig:7}
\end{figure}

We apply the double-focusing algorithm to the field data for this target. For this, we define 241 virtual sources and 241 virtual receivers with a spacing of $12.5$ m at $2500$ m depth extending from $9000$ m to $12000$ m over the upper boundary of the first target area. The up-going wavefield resulting from double-focusing is used as input for LSRTM and is called 'observed data' in the following. Figure~\ref{fig:6} shows the 'observed', and predicted data next to the residuals of Marchenko double-focusing target-oriented LSRTM. Moreover, Figure~\ref{fig:7} shows the same but for a conventional double-focusing approach. Conventional means using the inverse of the direct arrival between the target and the surface as the redatuming operator instead of the Marchenko focusing functions. The non-physical noise in the data is caused by imperfect surface multiple elimination in this part of the data. The computational advantage of target-oriented LSRTM with double-focused data is twofold. First, this algorithm reduces the spatial dimension of the problem, and second, it reduces the temporal dimension of the problem as well. The original recording time of the data at the surface is 8 s, whereas the temporal length of the double-focused data is 0.5 s.

\begin{figure}
    \centering
    \includegraphics[width=1\textwidth]{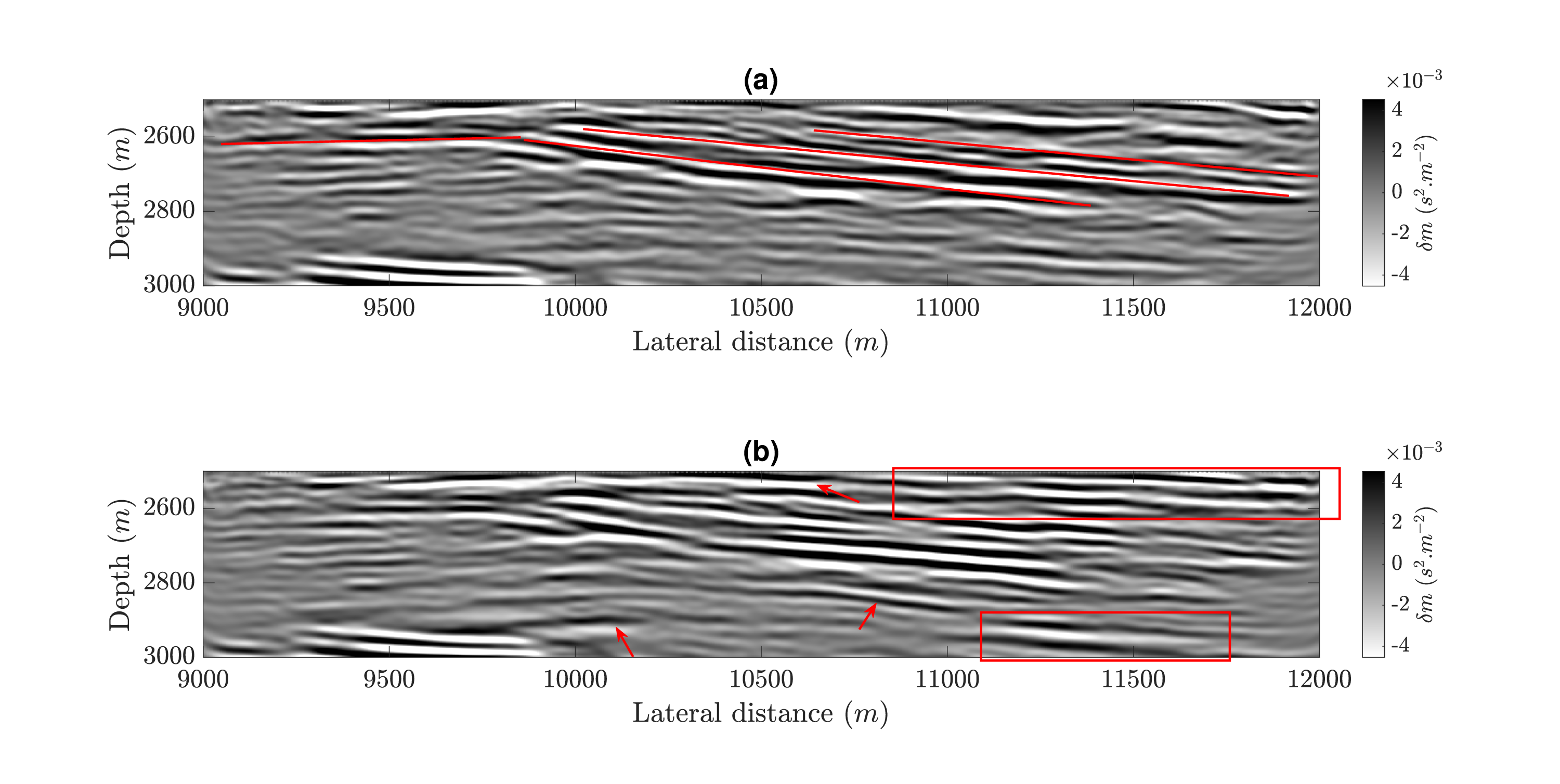}
    \caption{Comparison of images obtained with Marchenko target-oriented LSRTM (a) and Conventional target-oriented LSRTM (b). Red lines in panel (a) delineate some trends that are not visible in panel (b), and the red arrows and rectangles in panel (b) show some events that may be internal multiple reflection artifacts that are suppressed in panel (a).}
    \label{fig:8}
\end{figure}

 Figure~\ref{fig:8} compares the LSRTM images of using Marchenko and conventional double-focused data as input. Figure~\ref{fig:8} shows some improvements from using Marchenko double-focused wavefields compared to conventional double-focused ones. A comparison of our results with the results of ~\cite{EricDavy} and \cite{Ypma} confirms that the suppressed events are likely internal multiple reflections.

\begin{figure}
    \centering
    \includegraphics[width=1\textwidth]{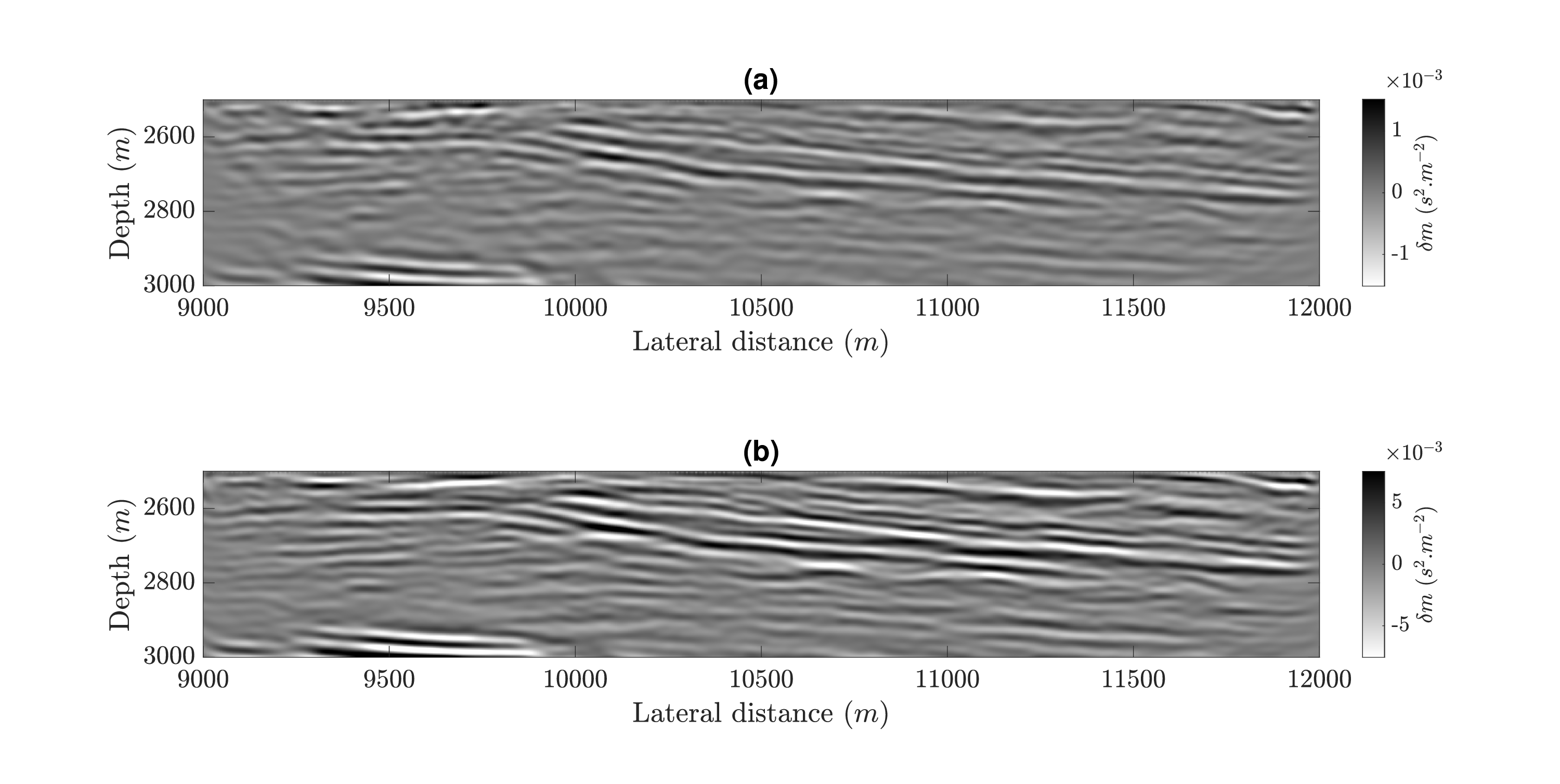}
    \caption{Comparison of images obtained with Marchenko target-oriented RTM (a) and LSRTM (b) of the first target.}
    \label{fig:9}
\end{figure}

Moreover, Figure~\ref{fig:9} compares the RTM and LSRTM images of Marchenko double-focused data as input. The LSRTM algorithm improved the quality of the image.

\subsubsection{Target of interest 2}

\begin{figure}
    \centering
    \includegraphics[width=1\textwidth]{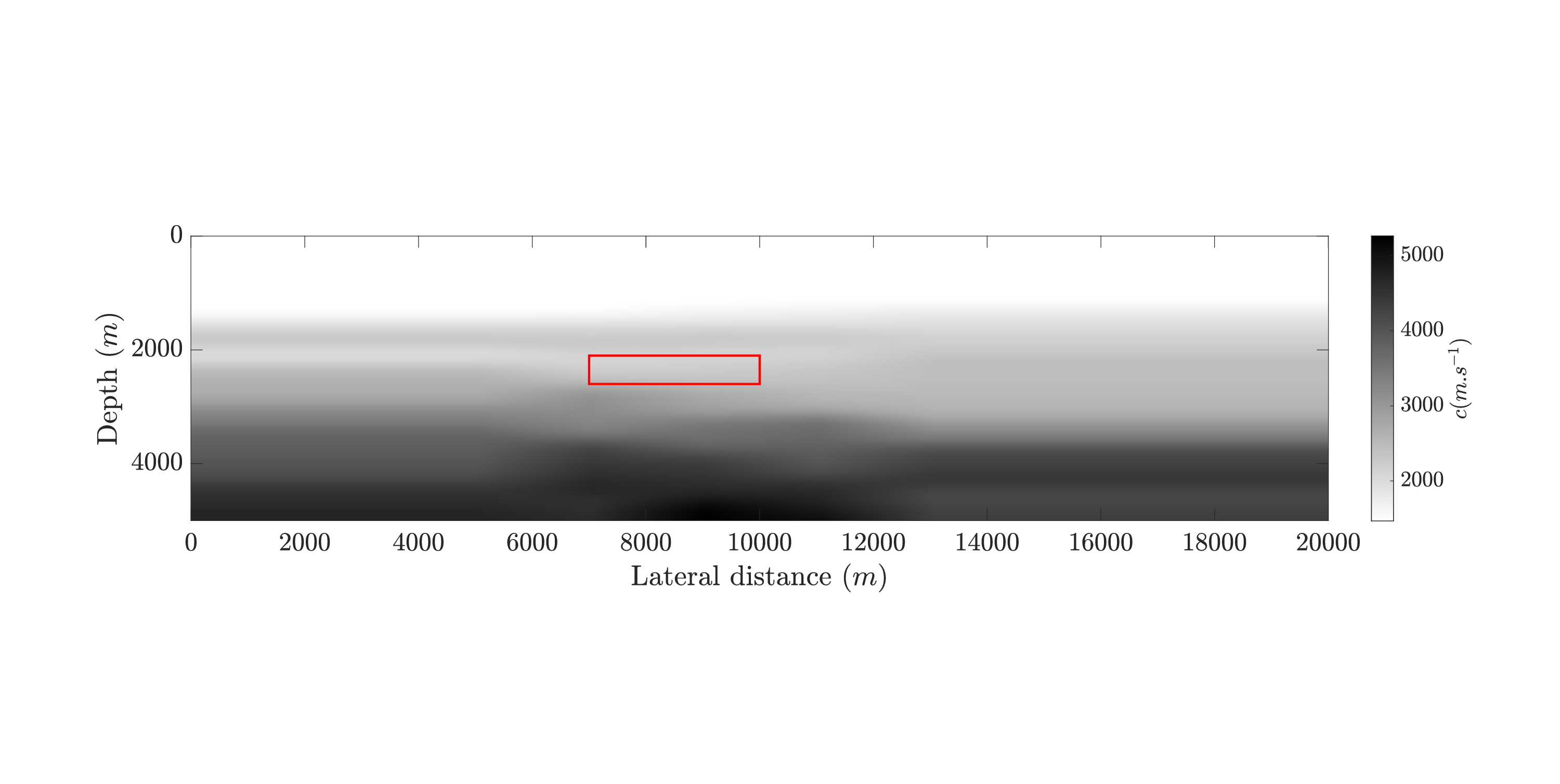}
    \caption{The smooth velocity model provided by Equinor for migration. The red rectangle inside the velocity model indicates the second target area and the virtual sources and receivers' positions are at the upper boundary.}
    \label{fig:10}
\end{figure}

Here, we choose another target. This target is located between depths of $2100$ m and $2600$ m and lateral extension from $7000$ m to $10000$ m, as shown in Figure~\ref{fig:10}. Virtual sources and receivers are located at the upper boundary of this target area.

\begin{figure}
    \centering
    \includegraphics[width=1\textwidth]{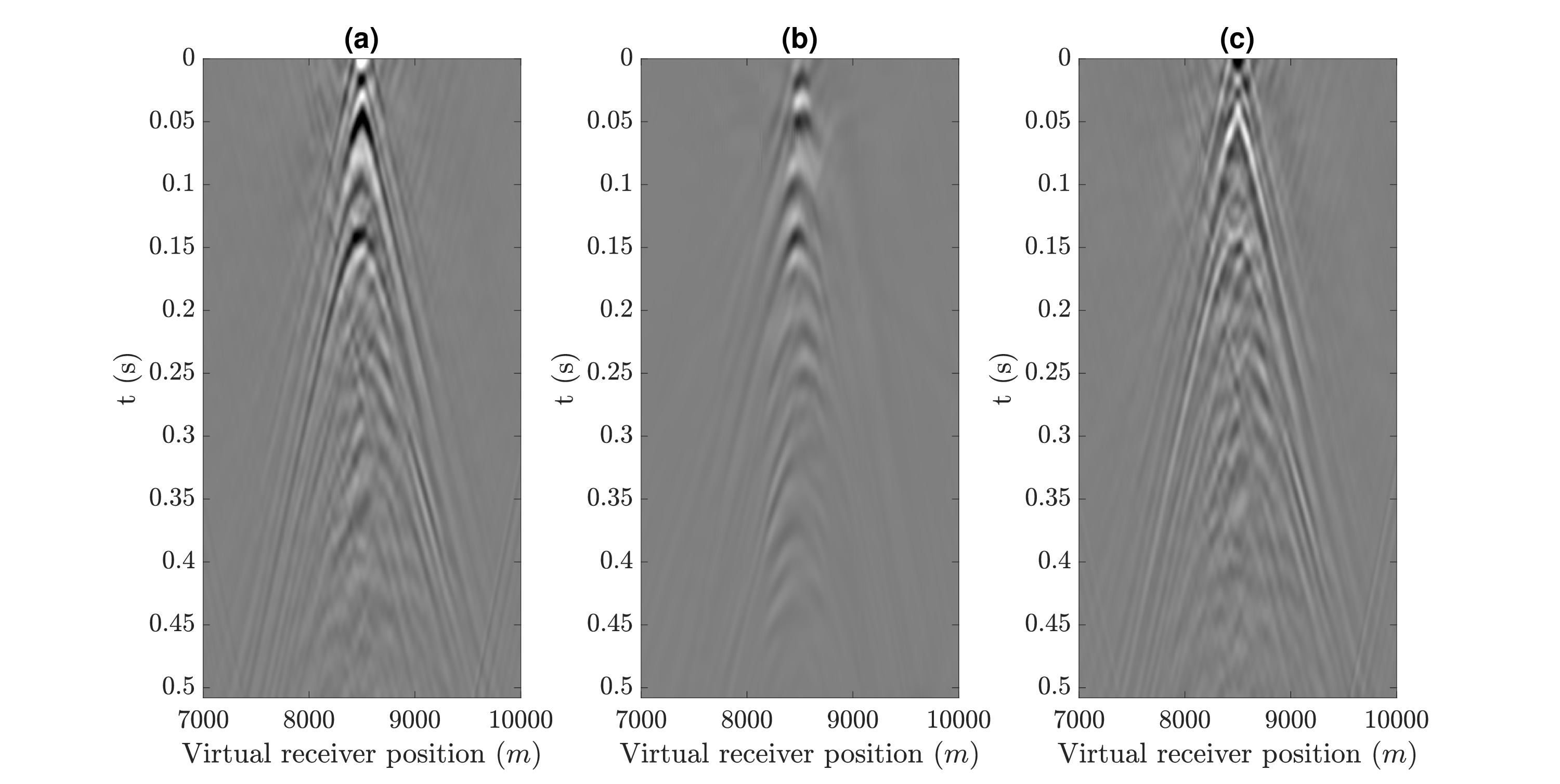}
    \caption{Marchenko double-focused data with a virtual source located at $\textbf{x}_{vs}=$ (8500 m, 2100 m) and virtual receivers at the same depth as virtual sources. a) 'observed data' obtained by Marchenko double focusing, b) predicted data after 35 iterations of LSRTM, and c) residuals after 35 iterations of LSRTM.}
    \label{fig:11}
\end{figure}

\begin{figure}
    \centering
    \includegraphics[width=1\textwidth]{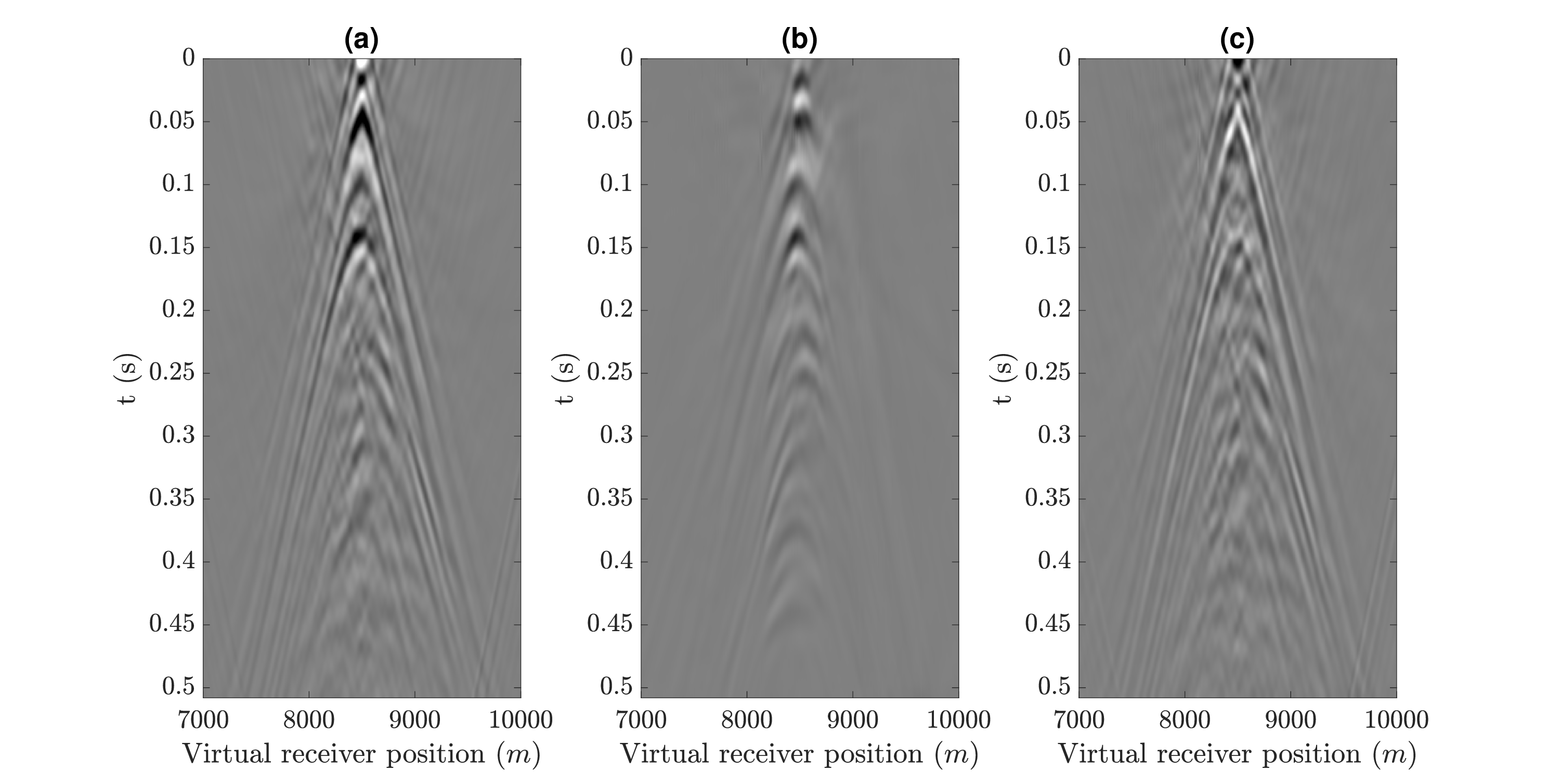}
    \caption{Conventional double-focused data with a virtual source located at $\textbf{x}_{vs}=$ (8500 m, 2100 m). a) 'observed data' obtained by conventional double focusing, b) predicted data after 35 iterations of LSRTM, and c) residuals after 35 iterations of LSRTM.}
    \label{fig:12}
\end{figure}

Figure~\ref{fig:11} shows the 'observed', and predicted data with corresponding residuals of the Marchenko double-focusing approach. Figure~\ref{fig:12} shows the same for the conventional double-focusing approach.

\begin{figure}
    \centering
    \includegraphics[width=1\textwidth]{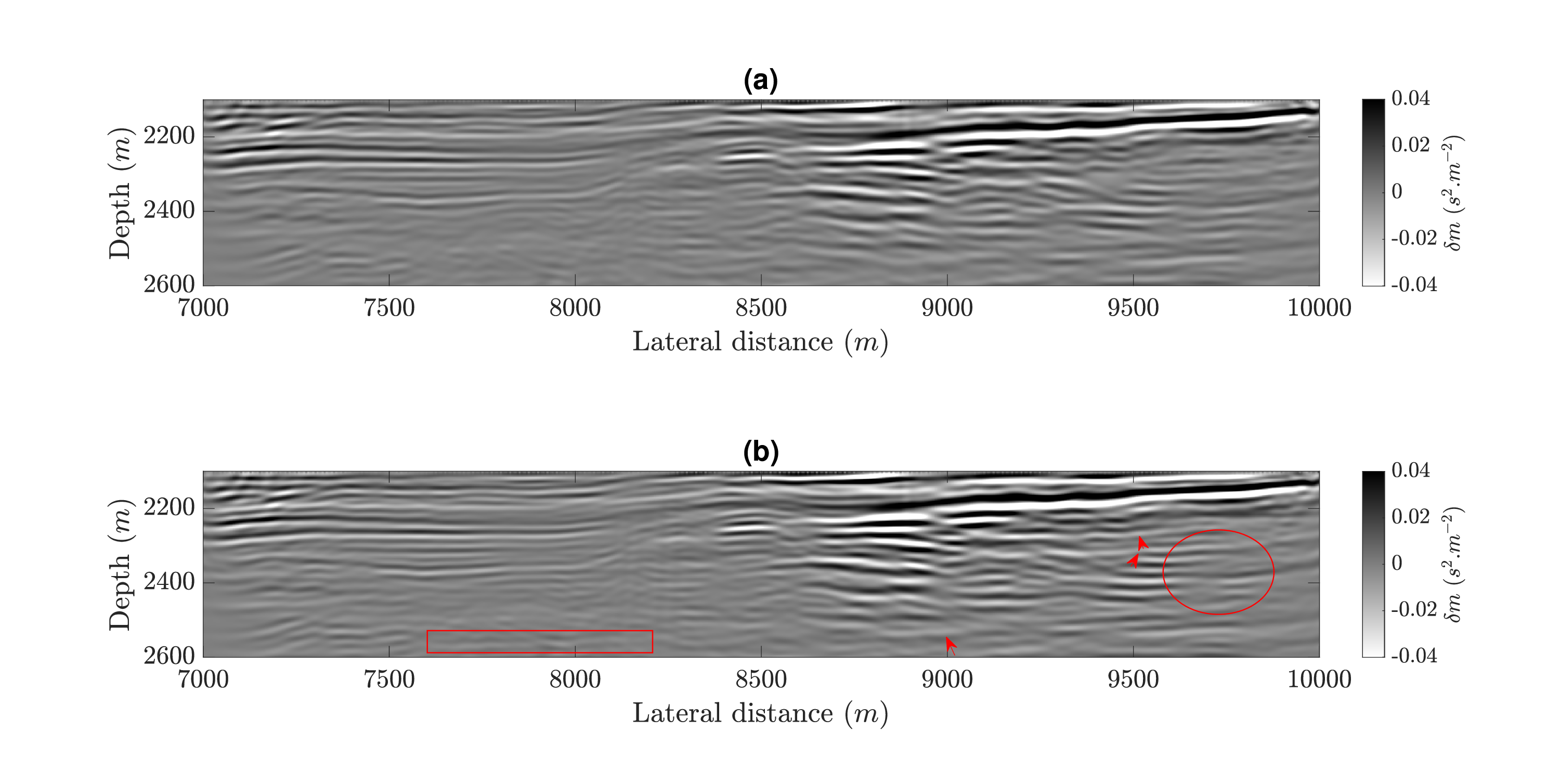}
    \caption{Comparison of images obtained with Marchenko target-oriented LSRTM (a) and Conventional target-oriented LSRTM (b). The red arrows, rectangle, and ellipse in panel (b) indicate some of the internal multiple reflection artifacts that are suppressed in panel (a).}
    \label{fig:13}
\end{figure}

\begin{figure}
    \centering
    \includegraphics[width=1\textwidth]{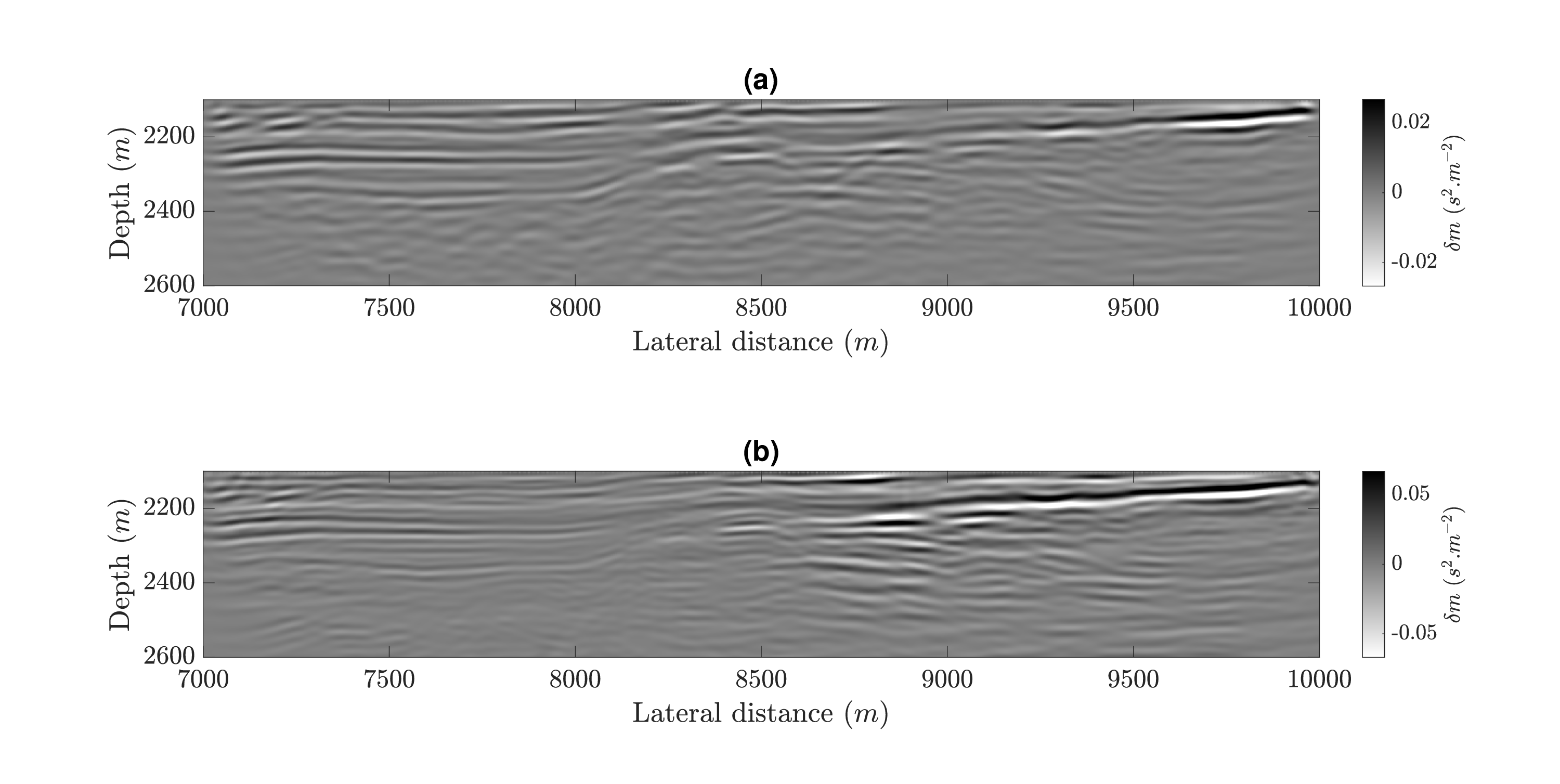}
    \caption{Comparison of images obtained with Marchenko target-oriented RTM (a) and LSRTM (b) of the second target.}
    \label{fig:14}
\end{figure}

Moreover, Figure~\ref{fig:13} shows the LSRTM images of the target-oriented algorithm with Marchenko and conventional double-focusing. The red arrows, the red rectangle, and the red ellipse indicate the internal multiple reflections that our method suppresses. Figure~\ref{fig:14} shows the RTM and LSRTM images of the target-oriented algorithm with Marchenko double-focusing. The LSRTM algorithm increases the quality and resolution of the image.

\section{Discussion}
In section 2 of this paper, we derive a target-oriented LSRTM algorithm based on double-focusing that can significantly reduce the dimensions of the problem, which also reduces the computational costs of the LSRTM algorithm. We also integrate the Marchenko double-focusing algorithm with our target-oriented LSRTM algorithm to reduce the artifacts caused by internal multiple reflections.

To demonstrate the advantages of our proposed algorithm, we applied it to a dataset acquired by Equinor in the Norwegian Sea in 1994. We chose two different target zones. Figures~\ref{fig:5} and ~\ref{fig:10} show our targets of interest embedded in the entire domain of the region. This spatial dimension reduction is to validate the first advantage we mentioned above. Figures \ref{fig:6}(a), \ref{fig:7}(a), \ref{fig:11}(a), and \ref{fig:12}(a) show the double-focused data with a recording duration of $0.5$ s, whereas the recording time of the original data is $8$ s.\\

To move forward with our investigation, we showed the imaging results with double-focusing for both targets. Figure~\ref{fig:8} compares the conventional and Marchenko double-focusing target-oriented LSRTM imaging results. The first panel (fig. \ref{fig:8}(a)) shows the LSRTM result of our proposed algorithm with Marchenko double-focused data, and the second panel (fig. \ref{fig:8}(b)) shows the LSRTM results with conventional double-focused data. Comparing these two panels reveals that using Marchenko double-focused wavefields leads to better visualization of true events and fewer artifacts due to internal multiples, delineated by the lines and arrows in those panels. Moreover, Figure~\ref{fig:9} shows the resolution and quality improvement of target-oriented LSRTM compared to target-oriented RTM with Marchenko double-focusing.\\

The same discussion holds for the second target. Figure~\ref{fig:13} shows a comparison between conventional and Marchenko double-focusing target-oriented LSRTM images where the internal multiple suppression is visible and indicated by arrows and an ellipse, and Figure~\ref{fig:14} shows the RTM and LSRTM images of Marchenko double-focusing. The quality and resolution of the image are increased noticeably. We use the internal multiple elimination results of \cite{EricDavy} and \cite{Ypma} as benchmarks for our results.

In both target areas, the double-focused gathers experience multiple flawed preprocessing stages that cannot be adequately explained by the forward modeling or the Marchenko approach. The non-physical artifacts within the data primarily arise from incomplete surface multiple removal and the impact of 3D effects in 2D datasets. These spurious elements in the data have led to a substantial residual.

\section{Conclusion}
This paper discusses a target-oriented LSRTM algorithm based on double-focusing. The advantages of this algorithm are: 1) reduction of the spatial dimensions of the problem by choosing a smaller target of interest, and 2) reduction of the temporal dimension of the problem by creating both virtual sources and receivers at the boundary of the target, which leads to lower computational costs. One can opt for sophisticated redatuming algorithms such as Marchenko redatuming and double-focusing to create virtual sources and receivers. The advantage of using Marchenko double-focusing compared to a more conventional redatuming algorithm is the ability to predict the internal multiple reflections inside the overburden and a reduction of artifacts due to these multiple reflections.

The reason for choosing double-focusing instead of multidimensional deconvolution (MDD) is to avoid another inversion step in our algorithm. MDD is an inversion process with its own uncertainties and associated errors. On the other hand, the double-focusing is a multidimensional convolution process with a stable output. As shown in \cite{Shoja3} and this paper, the predicted data, which uses the double-focused down-going wavefield at the boundary of the target, can predict the interactions between the target and the overburden and fit them to the double-focused observed data. 

Present-day seismic imaging and inversion applications need more accurate and higher-resolution images. Computing higher-resolution images demands significant amounts of computational power and time. Thus, devising algorithms that can reduce this computational burden is essential. Our proposed target-oriented algorithm is not only able to greatly reduce the spatial and temporal dimensions of the problem but can also reduce the artifacts due to internal multiple reflections by integrating Marchenko double-focusing with the LSRTM algorithm. Consequently, our algorithm enables us to create higher-resolution images with fewer artifacts at a lower computational cost.

\section{ACKNOWLEDGMENTS}
The authors would like to thank Eric Verschuur for fruitful discussions and feedback. This work was supported by the European Union’s Horizon 2020 Research and Innovation Program: European Research Council under Grant 742703.

\newpage

\bibliographystyle{seg}  
\bibliography{MyRefs}

\end{document}